# Chemical Elements Abundance in the Universe and the Origin of Life

Vlado Valkovic

Ruder Boskovic Institute, Zagreb, Croatia

Correspondence:

e-mail address: valkovic@irb.hr

Telephone: +385-1-468-0101, Fax: +385-1-468-0239

**Running title: Elements Abundance and Origin of Life**

**Abstract**

Element synthesis which started with p-p chain has resulted in several specific characteristics including lack of any stable isotope having atomic masses 5 or 8. The C/O ratio is fixed early by the chain of coincidences. These, remarkably fine-tuned, conditions are responsible for our own existence and indeed the existence of any carbon-based life in the Universe.

Chemical evolution of galaxies reflects in the changes of chemical composition of stars, interstellar gas and dust. The evolution of chemical element abundances in a galaxy provides a clock for galactic aging. On the other hand, the living matter on the planet Earth needs only some elements for its existence.

Compared with element requirements of living matter a hypothesis is put forward, by accepting the Anthropic Principle, which says: life as we know, (H-C-N-O) based, relying on the number of bulk and trace elements originated when two element abundance curves, living matter and galactic, coincided.

This coincidence occurring at particular redshift could indicates the phase of the Universe when the life originated. It is proposed to look into redshift region z = 0.5–2.5 (approximately t = −5.2x10$^9$ to -11.3x10$^9$ years) where many galaxies have been observed and to use these data to study the evolution of metallicity with respect to the other properties of galaxies in order to determine the time when universal element abundance curve coincided with the element abundance curve of LUCA. The characteristic properties of the latter have been transmitted by the genetic code while the universe element abundance curve changed as the galaxies aged.

## 1. Introduction

We start with the list of some properties of living matter that separate it from non-living matter: (i) Living matter is organized into complex structures based on organic molecules. (ii) Living matter maintains some type of homeostasis. (iii) Living matter grows and develops. (iv) Living matter reproduces and passes on genetic material as a blueprint for growth and subsequent reproduction. (v) Living matter acquires matter and energy from the external environment and converts it into different forms. (vi) Living matter responds to stimuli from the environment. (vii) Living matter evolves. (viii) Living matter is capable of taking the required chemical elements against the concentration gradient.

There are many proposed scenarios on the origins of life. Most of them, by considering above requirements, are actually concerned with the sustainability of life rather than the origin itself.

On other hand, the distribution of for life essential elements within the periodic table of elements might contain some clues about the origin of element requirements and possibly about the development of life in prebiotic time. The assumption that the vital body chemistry should bear similarity to the primordial chemical environment is a reasonable one. An organism would not make itself dependent on a rare element for its existence providing a more abundant element could play the same role. It is obvious that the presence of an element is a necessary prerequisite for the development of an essential metabolism based on that element (Valkovic 1990).

It was Hoyle (1953) who realized that this remarkable chain of coincidences – the unusual stability of berillium, the existence of an advantageous excitation level in $^{12}C$ and the non-existence of disavantageous level in $^{16}O$ – were necesarry, and remarkably fine-tuned, conditions for our own existence and indeed the existence of any carbon-based life in the Universe. For the ratio Hoyle deduced $^{12}C$ / $^{16}O$ = 1 / 3. Today, the solar C/O is 0.54.

The cosmic time scale for the evolution of life as proposed by Sharov and Gordon (2013) has important consequences: (1) life took a long time (ca. $5 \times 10^9$ years) to reach the complexity of bacteria; (2) the environments in which life originated and evolved to the prokaryote stage may have been quite different from those envisaged on Earth; (3) there was no intelligent life in our universe prior to the origin of Earth, thus Earth could not have been deliberately seeded with life by intelligent aliens; (4) Earth was seeded by panspermia; (5) experimental replication of the origin of life from scratch may have to emulate many cumulative rare events; and (6) the Drake equation for estimating the number of civilizations in the universe is likely wrong, as intelligent life has just begun appearing in our universe.

The approach to the problem of the origin of life could be as follows. By accepting the Strong Anthropic Principle which says "The Universe must have those properties which allow life to develop within it at some stage in its history" we can put forward a hipothesis:

Life as we know, (H-C-N-O) based, relying on the number of bulk (Na-Mg-P-S-Cl-K-Ca) and trace elements (Cr-Mn-Fe-Co-Ni-Cu-Zn-Se-Mo-I-W, and possibly Li-B-F-Si-V-As) originated when two element abundance curves coincided. The two element abundance curves are the element abundance in living matter and the universal element abundance curves.

It is clear that all organisms known on the tree of life should share a common ancestor. One of the most important challenges has been determining the nature of the Last Universal Common Ancestor (LUCA). Hyperthermophilic organisms closest to the root suggest that LUCA was hyperthermophilic. Efforts are in progress to evaluate the possible elemental composition of LUCA, see (Chopra and Lineweaver 2015). LUCA almost certainly had a simple metabolism compared to modern forms, with less than a full complement of essential amino acids and perhaps a few non-specific enzymes. Aromatic amino acids appear to have arisen some time after LUCA arose, as did more complex and specific enzymes. In addition to the elemental composition LUCA should be able to survive the intergalactic transfer which would then support the idea of cosmic dust particles as its habitat.

In order to gain an insight about $z_{o.l.}$ we propose to look into redshift region z = 0.5–2.5 (approximately t = −5.2x$10^9$ to -11.3x$10^9$ years) where many galaxies have been observed and to use these data to study the evolution of metallicity with respect to the other properties of galaxies. The abundance ratios: $(C:O)_{\Delta z}/(C:O)_{life}$; $\Sigma(H:C:N:O)_{\Delta z}/\Sigma(H:C:N:O)_{life}$; $\Sigma(Na: Mg: P: S: Cl: K: Ca)_{\Delta z}/\Sigma(Na: Mg: P: S: Cl: K: Ca)_{life}$; $\Sigma(Cr: Mn: Fe: Co: Ni: Cu: Zn: Se: Mo: I: W)_{\Delta z}/\Sigma(Cr: Mn: Fe: Co: Ni: Cu: Zn: Se: Mo: I: W)_{life}$ and $\Sigma(Li: B: F: Si: V: As)_{\Delta z}/\Sigma(Li: B: F: Si: V: As)_{life}$ should be calculated for several $\Delta z$ intervals of z-value. The parameter measuring the closenest of two abundance curves, $C_{AC}$, should be defined on the most appropriate manner.

## 2. Nucleosynthesis

It is accepted believe that the element synthesis in Universe starts with the p-p chain with the final result being $^4$He, helium. The helium that is produced as the "ash" in this thermonuclear hydrogen "burning" cannot undergo fusion reactions at these temperatures or even substantially above because of a basic fact of nuclear physics in our Universe: there are no stable isotopes (of any element) having atomic masses 5 or 8. Non-existence of nuclei of atomic weight 5 and 8

could be overcome with the resonance $\leq 400$ keV in $^7$Li in the reaction: $^3$H+$^4$He$\rightarrow$$^7$Li+$\gamma$ (Alpher and Herman 1950). No such resonance was found to exist and attempts to bridge the gap at mass 5 were abandoned. Only at very high temperatures, of order $10^8$ K, can this bottleneck be circumvented by a highly improbable reaction. At those temperatures, the fusion of two $^4$He nuclei forms highly unstable $^8$Be at a fast enough rate that there is always a very small equilibrium concentration of $^8$Be at any one instant (Salpeter 1952).

This small concentration of $^8$Be can begin to undergo reactions with other $^4$He nuclei to produce an excited state of the mass-12 isotope of carbon. This excited state is unstable, but a few of these excited carbon nuclei emit a gamma-ray quickly enough to become stable before they disintegrate. This extremely improbable sequence is called the triple-alpha process because the net effect is to combine 3 alpha particles (that is, 3 x $^4$He nuclei) to form a $^{12}$C nucleus.

Most of the carbon produced in the core of a massive star will still be present when the star explodes as a supernova and spreads its material into interstellar space. $^{12}$C+$\alpha$ masses are 7.1695 MeV so the reaction $^{12}$C($\alpha,\gamma$)$^{16}$O must occur through a known nonresonant level at 7.1169 MeV. Since the 7.1169 level is just below $^{12}$C+$\alpha$=7.1695 MeV, resonance cannot occur. With 7.65 MeV resonance the carbon yield has increased by a factor of about $10^7$ compared to $3\alpha\rightarrow$$^{12}$C.

Although at lower temperatures the p-p chain dominates, with rising temperatures there is a sudden transition to dominance by the CNO cycle (Weizsäcker 1938, Bethe 1939), which has an energy production rate that varies strongly with temperature.

The fusion of hydrogen to helium by either the p-p chain or the CNO cycle requires temperatures of the order of $10^7$ K or higher, since only at those temperatures will there be enough hydrogen ions in the plasma with high enough velocities to tunnel through the Coulomb barrier at sufficient rates.

If a star had sufficient mass, though, eventually enough C would accumulate so that the temperature and density reach a point where C nuclei could be fused into neon nuclei. This carbon burning core would be surrounded by two outer shells, the innermost burning He, and the outermost burning H. This pattern of the central core collapsing and increasing temperature continues until a further round of fusion occurs and more shells form. As the fusion process continues the concentration of Fe increases in the core of the star, the core contracts, and the temperature increases to a point where Fe can undergo nuclear reactions.

Fe nuclei are the most stable of all atomic nuclei. Because of this, when they undergo nuclear reactions, they don't release energy, but absorb it. Therefore, there is no release of energy to balance the force of gravity. In fact, there is actually a decrease in the internal pressure that works with gravity to make the collapse of the core more intense. In this collapse, the Fe nuclei in the central portion of the core are broken down into alpha particles, protons, and neutrons and are compressed even further.

Eventually, the outer layers of material rebound off the compressed core and are thrown outward. This sets the stage for a tremendous collision between the recoiling core layers and the collapsing outermost layers. Under the extreme conditions of this collision, two things happen that lead to the formation of the heaviest elements. First, the temperature reaches levels that cannot be attained by even the most massive stars. This gives the nuclei present large kinetic energies, making them very reactive. Second, because of the breaking apart of the iron nuclei in the central core, there is a high concentration of neutrons that are ejected from the core during the supernova. These neutrons are captured by surrounding nuclei, and then decay to a proton by emitting an electron and an antineutrino. With the large neutron flux created during a supernova, this neutron capture/decay sequence can be repeated many times, adding protons to form increasingly more massive nuclei. These conditions exist for only a short time, but long enough to form the highest mass nuclei. Because of this "rebound explosion," all the outer layers of the star, enriched with the higher mass nuclei, are blown off into space, and this material will later make its way into other nebulas to become incorporated into other stars (where the same cycle of events will be repeated). Each cycle uses up more of the H and He from the early universe and creates greater amounts of the higher mass elements.

Final abundances of elements are determined in supernova explosions. Larger quantities of lighter elements in the present universe are therefore thought to have been restored through billions of years of cosmic ray (mostly high-energy proton) induced breakup of heavier elements in interstellar gas and dust. Table 1 lists the basic nuclear reaction links.

With the estimates of the production ratio of $U^{235}/U^{238}$ in the r-process, Burbidge et al. (1957) were able to calculate the age of these uranium isotopes (and by inference of the other r-process isotopes) using their presently observed terrestrial abundance ratio. Their best value for the time of a single-event synthesis of $U^{235}$ and $U^{238}$ is thus $6.6 \times 10^9$ years ago. If these isotopes were produced in a single supernova, this is their date. For each change of a factor of two in the production ratio of the uranium isotopes, this figure changes by $0.85 \times 10^9$ years.

## 3. Element Abundances

Our solar system is the result of the gravitational collapse of a small part of a giant molecular cloud. It is often assumed that the Sun, the planets, and all other objects in the solar system are formed from a hot gaseous nebula with well defined chemical and isotopic composition. The findings of comparatively large and widespread variations in oxygen isotopic composition has cast some doubt upon this assumption. Additional evidence of incomplete mixing and small scale homogenization of elements and nuclides in the primordial solar nebula is provided by the detection of huge (up to the percent level) isotope anomalies of some heavy elements in meteoritic silicon carbide (SiC) grains.

Solar system abundances are quite similar to those found in most stars and interstellar material in our neighborhood and in corresponding parts of other galaxies where, however, minor variations (within a factor of 3 or so either way) may occur in the relative amounts of hydrogen and helium, on one hand, and carbon and heavier elements on the other. This reflects the fact that hydrogen and the bulk of helium are relics from the 'Big Bang', whereas heavier elements (and a minority of the helium) result from nuclear reactions in stars or in the interstellar medium.

Carbon and heavier elements tend to be relatively more abundant in the central regions of large galaxies (such as our own) than in their outer parts or in small galaxies; in stars belonging to the outer spheroid halo of our Galaxy carbon and heavier elements may be deficient by factors of up to 1 000 or more (relative to hydrogen and helium) when compared to Solar system values, and among these elements, carbon, nitrogen, iron and elements such as barium (resulting from the 'slow' neutron capture or s-process in the progenitor stars) can be deficient by larger factors than oxygen, magnesium and other 'α-particle' elements synthesized in massive stars which undergo supernova explosions after 10 million years or so. Peculiar over- and under-abundances of various elements can also be found in some highly evolved stars as a result of internal nuclear reactions and in the surface layers of certain stars where diffusive separation of elements seems to have occurred.

Beyond the region of nuclides with mass numbers of 56 (the "Fe-peak" region), abundances decline more or less smoothly and spike at certain mass number regions. The nuclides beyond the Fe peak are products from neutron capture processes. The peaks in the distribution correspond to regions where either nuclides are preferentially made by the slow-neutron capture (s-) process operating in red giant stars (e.g, Y and Ba regions) or by the rapid-neutron capture (r-) process probably operating in supernovae, e.g., Pt region; see (Wallerstein et al. 1997, Woosley et al. 2002, Sneden et al. 2008) for reviews on stellar nucleosynthesis. Here the "slow" and "rapid" are in reference to beta-decay timescales of the intermediate, unstable nuclei produced during the neutron capture processes. The nuclide yields from these processes depend on the neutron energies and flux, but also on the abundance and stability of the target nuclei against neutron capture which in turn depends on Z and N. Hence the abundance distribution becomes controlled by the more stable "magic" nuclides that serve as bottlenecks for the overall yields in the neutron capture processes.

Some 2000 exoplanets, or planets outside the solar system, have been discovered to date. Until recently, infrared observations of exoplanetary atmospheres have typically been interpreted using models that assumed solar elemental abundances. However, recent observations have revealed deviations from predictions based on such classification schemes, and chemical compositions retrieved from some data sets have also indicated non-solar abundances, i.e. deviation in C/O ratio from 0.5, see for example Marboeuf et al. (2014).

For example, the carbon-to-oxygen ratio in a planet provides critical information about its primordial origins and subsequent evolution. A primordial C/O greater than 0.8 causes a carbide-dominated interior, as opposed to the silicate-dominated composition found on Earth; the atmosphere can also differ from those in the Solar system.

Present day solar system composition is discussed in works (Lodders et al. 2009; Lodders 2010). Abundance peaks at mass numbers for closed proton and neutron "shells". The "magic numbers" for nuclear stability are 2, 8, 20, 28, 50, 82, and 126; and nuclides with Z and/or N equal to these magic numbers are the ones that show large abundances in the diagram of abundance versus mass number (A=Z+N). This is particularly notable for the light doubly-magic nuclei with equal magic Z and N, e.g., $^4$He (Z=N=2), $^{16}$O(Z=N=8), and $^{40}$Ca (Z=N=20).

Solar system element abundances $4.56 \times 10^9$ years ago are discussed by (Lodders et al. 2009; Lodders 2010): The present day photospheric abundances described above are different from those in the Sun $4.56 \times 10^9$ years ago, at the beginning of the solar system. Two processes affected the solar abundances over time. The first is element settling from the solar photosphere into the Sun's interior; the second is decay of radioactive isotopes that contribute to the overall atomic abundance of an element.

The facts are following: The curve representing element abundances is changing in the time. The changes can be observed on the scale $10^9$ years. This is true not only for the solar system but also for the rest of our galaxy (stars and interstellar matter) and for the Universe as whole. The evolution of chemical element abundances in a galaxy provides a clock for galactic aging. There is a relation between the ages and metal abundances of stars: on average, older stars contain less iron than younger stars.

It is in principle possible to find the chemical composition of a galaxy as a function of position and time by measuring abundances of stars with different birthplaces and ages, provided that their atmospheres represent the composition of the gas from which they were formed. The chemical abundance of the gas and dust particles in a star-forming galaxy is evolving with time. The metal abundance of the gas and dust particles, and of subsequent generations of stars, is increasing with time.

Of interest are especially elements in the Fe-peak of abundance curve. The Fe-peak elements are created in various late burning stages in stars, as well as in supernovae. In addition, the [Cr/Zn] ratio is very suitable for such investigations. From the earliest abundance measurements it was established that this ratio in stars is generally sub-solar, as expected if a fraction of the Cr has been incorporated into dust grains (Pettini 2003).

## 4. Organic cycle / Living matter

The living matter is capable of taking the required chemical elements against the concentration gradient. In addition, the autopoiesis is the necessary condition shared by any living entity and therefore is universal (Lucantoni and Luisi 2012).

The organic cycle includes the eleven elements: H, C, N, O, Na, Mg, P, S, Cl, K and Ca which form the bulk of the living matter. They all have very low atomic weights and they belong to the lowest 20 elements of the periodic table. These elements are also the most abundant elements in the Universe. The only nonessential light elements, with exclusion of noble gases, are Li, Be, B (boron is essential for some plants) and may-be Al. In addition, a number of elements have been recognized as essential trace elements. For example, essential trace elements for warm-blooded animals are: F, Si, V, Cr, Mo, Fe, Co, Ni, Cu, Zn, Se, Mo, Sn, I and W. For the most part the essential trace elements are transition metals with unfilled d-orbital. None of the heavy elements beyond tungsten (Z=74) has ever been shown to have any physiological significance.

Distribution of essential chemical elements for mammals is shown in Table 2 and Fig. 1. The chemical elements can be grouped as: (i) Four basic elements: H, C, N, O. (ii) Essential bulk elements: Na, Mg, P, S, Cl, K, Ca. (iii) Essential trace elements: Cr, Mn, Fe, Co, Ni, Cu, Zn, Se, Mo, I, W. (iv) No confirmed functions in humans, but plants, animals, etc: Li, B, F, Si, V, As. (v) Rest: Environmental pollutants.

Figure 1 shows the concentration values of chemical elements in living matter: groups (i) and (ii) are marked by red dots, group (iii) green dots, group (iv) yellow dots and group (v) black dots.

## 4.1 Origin of trace element requirements

Research on trace elements in chemical evolution should be reviewed from three points of view. They are (i) the origin of the essentiality of trace elements in present biological systems; (ii) the possible roles of trace elements in chemical evolution; and (iii) the origin of enzymatic activity with metal ions, i.e., the origin of metalloenzymes. The role of trace elements in chemical evolution has been increasingly considered as the role of trace elements in the present biological systems are being documented (Kobayashi and Ponnamperuma 1985). Origin of trace element requirement by living matter is discussed by (Valkovic 1990).

In explaining why a particular element has been selected for an essential biochemical role, one of the possible approaches is as follows: If an element has not been selected, this may be because its abundance in the available environment is too low, either on the absolute scale, or in comparison with some other element that can play the same role.

Groups of elements required for the life are:
1. Unique requirement dating from the origin of life: H, C, N, O, K, Mg, P, S, Fe.
2. Unique requirement may-be acquired later: B, Se, I.
3. Primordial requirement which could be satisfied by a number of elements – evolutionary adoption being made to the more abundant member: K *vs* Rb, Mg *vs* Be, S *vs* Se, Cl *vs* Br, H *vs* F.
4. The same as under 3. but a later acquisition: Ca *vs* Sr, Na *vs* Li, Si *vs* Ge.

The suggestion that the concentrations in seawater were those crucial to the origin of elemental requirements has been often discussed (see Chappell et al., 1974; Jukes, 1974; Orgel, 1974; Banin and Navrot, 1975). The ocean is thought to be in a steady state with sediments and the atmosphere. Although it is often remarked that the composition of organisms resembles that of the ocean, one might get different conclusions from careful examination of element concentration values in the seawater. High concentrations of nonessentials: Li, Br, Sr, Rb, and some others are evident. Metals which serve as essential trace elements are very strongly depleted in the ocean compared to plant or animal requirements. In addition, the content of both carbon and nitrogen is low compared to living matter concentrations and cosmic abundances.

The answer to the question of which environment should be compared to the composition of living organisms is that with which the organism has an intimate contact. Contrary to all expectations universal (cosmic) abundance curve of elements is in the best agreement with the distribution of essential elements within the periodic table. Essential elements are the most abundant elements; trace elements are almost all grouped in the secondary peak (around Fe). This is the region of the maximum nucleon binding energy in the nucleus, the fact which is responsible for the peak in the universal abundance curve.

In order to estimate the closeness of different abundance curves, one should evaluate relations between elements and/or groups of elements (C/O); (H,C,N,O); (Na, Mg, P, S, Cl, K, Ca); (Cr, Mn, Fe, Co, Ni, Cu, Zn, Se, Mo, I, W); (Li, B, F, Si, V, As) required for life processes and from well determined abundance curves for different surroundings (Universe, Solar system, planet Earth, crust, seawater) using data for example from (Emsley, J. 1998; Emsley, J. 2001), http://www.seafriends.org.nz/oceano/seawater.htm, and http://periodictable.com/Properties/A/UniverseAbundance.html or some other source. Table 3. shows abundance ratios for life vs. different media of interest.

## 5. Time of the origin of life

According to (Sharov 2012; Sharov and Gordon 2013) an extrapolation of the genetic complexity of organisms to earlier times suggests that life began before the Earth was formed. Life may have started from systems with single heritable elements that are functionally equivalent to a nucleotide. The genetic complexity, roughly measured by the number of non-redundant functional nucleotides, is expected to have grown exponentially due to several positive feedback factors: (1) gene cooperation, (2) duplication of genes with their subsequent specialization (e.g., via expanding differentiation trees in multicellular organisms), and (3) emergence of novel functional niches associated with existing genes.

Linear regression of genetic complexity (on a semi log scale), as measured by the length of functional non-redundant DNA per genome counted by nucleotide base pairs for mammals, fish, worms eukaryotes and prokaryotes, extrapolated back to just one base pair suggests the time of the origin of life to be $(9.7\pm2.5)x10^9$ years ago (Sharov 2006; Sharov 2012; Sharov and Gordon 2013). Time is counted backwards in $10^9$ of years before the present (time 0). This is shown in Table 4. showing a schematic view of the development of the universe since the Big Bang. It shows the proposed estimate for the origin of life, $(9.7\pm2.5)x10^9$ years ago. Note that the "Dark Ages" may have ended at -$13.55x10^9$ years (Zheng et al., 2012), with the Big Bang at -$13.75x10^9$ years (Jarosik et al. 2011).

There are other scenarios, according to McCabe and Lucas (2010), the origin of life in the Galaxy, or more specifically the start of Galactic habitability, is taken to have occurred $(8.5\pm5)x10^9$ years ago, where the central value of $8.5x10^9$ years reflects the work of Lineweaver et al (2004), Prantzos (2008) and Mattson (2009) and the range of uncertainty allows for the most extreme conceivable values from the first stars in the Universe to the origin of life on Earth.

If life originated long before the origin of Earth, we have to assume that Earth was contaminated with bacterial spores in process called panspermia, see also (Wallis and Wickramasinghe 2004; Schroeder 2015). The role of cosmic dust mineral grains in this process should be considered.

Houdek and Gough (2011) have attempted a seismic calibration of standard solar models with a view to improving earlier estimates of the main-sequence age and the initial heavy-element abundance. Their long-term goal has been to achieve a precision which could distinguish between planet formation occurring simultaneously with or subsequent to the formation of the Sun. Their current best estimates, around $(4.60\pm0.04)\times10^9$ years, the age is close to the previous preferred values – in particular, the age adopted for Christensen-Dalsgaard's Model S – and the implied present-day surface heavy-element abundance lies between the modern spectroscopic values quoted by Asplund et al. (2009) and Caffau et al. (2009).

A number of properties of the universe which are, according to physics as we currently understand it, completely arbitrary (that is, not determined by or calculated from any known law of physics, for example the ratio of the masses of the proton and electron, or the relative strengths of the four fundamental forces) appear to have values which, if changed only very slightly, would preclude the existence of beings like ourselves or, more speculatively, any intelligent life form whatsoever.

This apparent fine-tuning of the fundamental constants of nature is disturbing to many scientists, since it can be interpreted as evidence the universe was designed just for us--that the universe has a purpose, and that we are it. This seems to hark back to a prescientific world view, which all the evidence of hundreds of years of science has refuted point-by-point.

Recent advances in string theory and inflationary cosmology have led to a surge of interest in the possible existence of an ensemble of cosmic regions, or "universes," among the members of which key physical parameters, such as the masses of elementary particles and the coupling constants, might assume different values. The observed values in our cosmic region are then attributed to an observer selection effect (the so-called anthropic principle). The assemblage of universes has been dubbed "the multiverse", see (Davies 2004).

One way to explain this fine-tuning is to invoke the Anthropic Principle (Barrow and Tipler 1988) which, in its weakest form states that "We observe the universe with properties which permit us to exist because otherwise we wouldn't be here to do the observing." The Anthropic principle cannot be refuted--it is a logical tautology that one cannot observe conditions in the universe which preclude one's own existence. But invoking the Anthropic principle is, in a way, almost as unsettling as arguing that the values of the physical constants were preset by a benign creator so that we could eventually inhabit the universe. Either explanation essentially removes the subject matter, in this case the values of the physical constants, from the domain of science for in neither case would we expect to ever be able to calculate the values of the constants from first principles.

In 1953 Fred Hoyle realized that to make enough carbon inside the stars, there had to exist a resonance state of the carbon-12 nucleus at 7.68 MeV, which at the time was not known experimentally. Hoyle said "since we exist, then carbon must have an energy level at 7.6 MeV" – anthropic prediction! Strong Anthropic Principle (SAP) says: The Universe must have those properties which allow life to develop within it at some stage in its history (Carter 1974). Some authors (see Kragh 2010) argue that the excited levels in $^{12}$C and other atomic nuclei can be used as an argument for the predictive power of the anthropic principle.

## 6. Discussion and conclusions

In spite of many questions still awaiting for the answer, we can conclude that the distribution of for life essential elements holds the best possible clue on the origin of life questions.

Interstellar molecular clouds and circumstellar envelopes are factories of complex molecular synthesis (Ehrenfreund and Charnley 2000; Kwok 2004; van Dishoeck and Blake 1998). In addition to gas, interstellar material contains also small micron-sized particles. Gas-phase and gas-grain interactions lead to the formation of complex molecules. Surface catalysis on solid interstellar particles enables molecule formation and chemical pathways that cannot proceed in the gas phase only owing to reaction barriers. A surprisingly high number of molecules that are used in contemporary biochemistry on the Earth is already found in the interstellar medium, planetary atmospheres and surfaces, comets, asteroids and meteorites and interplanetary dust particles. Large quantities of extra-terrestrial material could be delivered via comets and asteroids to young planetary surfaces during the heavy bombardment phase.

The origin of dust particles in galaxies is still an unresolved matter (Gall et al. 2011; Matsura et al. 2009; Draine 2009; Dunne et al. 2011). The majority of the refractory elements are produced in supernova explosions, but it is unclear how and where dust grains condense and grow, and how they avoid destruction in the harsh environments of star forming galaxies. The recent detection of large mass of dust (0.5 $M_\odot$) in nearby supernova remnants (Matsura et al. 2011; Indebetouw et al. 2014; Gomez et al. 2012) suggests in situ dust formation, while others (Gall et al. 2014) reported rapid (40-240 days) formation of dust grains in the luminous supernova 2010jl. The detailed investigation of this dust reveals the presence of very large (exceeding one micrometre) grains, which resist destruction.

By now it is established that the curve representing chemical element abundances in any part of Universe is changing in time. Galaxies at different distances (different z-value) have different metallicity and different relative abundances of Fe-peak elements. However, there are many difficulties in determining element abundances of distant objects caused mainly by dust particles in the interstellar medium.

Some of the works discussing progress in this field are presented. For example, Mannucci et al. (2010) extracted from the literature three samples of galaxies at intermediate redshifts, for a total of 182 objects, having published values of emission-line fluxes, $M_*$, and dust extinction: $0.5 < z < 0.9$, $1.0 < z < 1.6$ and $2.0 < z < 2.5$. Roediger et al. (2013) presented an extenstive literature compilation of age, metallicity, and chemical abundance pattern information for the 41 Galactic globular clusters (GGCs) studied by Schiavon et al. (2005). Their compilation constitutes a notable improvement over previous similar work, particularly in terms of chemical abundances.

By comparison of element requirements of living matter with cosmic, solar, earth crust and seawater abundance curves, we have put forward a hypothesis, which says: life as we know, (H-C-N-O) based and relying on the number of bulk and trace elements originated on the dust particles at $z = z_{o.l.}$ when living matter and cosmic element abundance curves coincided.

The primordial process that turns enormous clouds of cosmic dust into newborn planets over millions of years has been recently observed directly. A protoplanet in the making around a young star ($2x10^6$ year old), LkCa15, in the neighborhood of Taurus, 450 light years from Earth, has been spotted (Sallum et al. 2015). So far nearly 2,000 exoplanets have been discovered and confirmed (Akeson et al. 2013). These could be locations where life might sustain and evolve providing appropriate duration of habitability conditions.

**References**

Alpher, R. And Herman, R. 1950. Theory of the origin and relative abundance distribution of the elements. Reviews of Modern Physics 22:153-212.

Asplund M., Grevesse N., Sauval A. J., Scott P. 2009. The Chemical Composition of the Sun. Annual Review of Astronomy & Astrophysics, vol. 47(1): 481-522.

Banin, A., and Navrot, J. 1975. Origin of life: clues from relations between chemical compositions of living organisms and natural environments. Science 189:550-551.

Barrow, J. D. and Tipler, F. J. 1988. The Anthropic Cosmological Principle. Oxford University Press, Oxford. ISBN 0-19-282147-4.

Bethe, H. A. 1939. Energy Production in Stars. Physical Review **55** (5): 434–456.

Burbidge, E. M., Burbidge, G. R., Fowler, W. A., Hoyle, F. 1957. Synthesis of Elements in Stars. Rev. Mod. Phys. 29(4):547-654.

Caffau E., Maiorca E., Bonifacio P., Faraggiana R., Steffen M., Ludwig H.-G., Kamp I., Busso M., 2009, The solar photospheric nitrogen abundance - Analysis of atomic transitions with 3D and 1D model atmospheres. Astronomy&Astrophysics 498: 877-884.

Carter, B. 1974. Large number coincidences and the anthropic principle in cosmology. Confrontation of Cosmological Theories with Observational Data (Longair, M. S., ed.)Riedel, Dordrecht, pp. 291-298.

Chappell, W. R., Meglen, R. R., and Runnells, D. D., 1974. Comments on "directed panspermia". Icarus 21: 513-515.

Chopra, A., Lineweaver, C. H., Brocks, J. J. and Ireland, T. R. 2010. Palaeoecophylostoichiometrics: Searching for the Elemental Composition of the Last Universal Common Ancestor.  Australian Space Science Conference Series: 9[th] Conference Proceedings, Sidney, 28-30 September 2009 (DV). National Space Society of Australia Ltd, ISBN 13: 978-0-9775740.  (www.tinyurl.com/ACetal10).

Christensen-Dalsgaard J. 2009. in Mamajek E. E., Soderblom D. R., Wyse R. F. G., eds, Proc. IAU Symp. 258, The Ages of Stars. Cambridge Univ. Press, Cambridge, p. 431.

Christensen-Dalsgaard, J., Däppen, W., Ajukov, S. V. et al. 1996. The Current State of Solar Modeling. Science 272: 1286-1292.

Davies, P. C. W. 2004. Multiverse Cosmological Models. Mod. Phys. Lett. A. 19(10): 727-743.

Ehrenfreund, P. and  Charnley, S. B. 2000. Organic molecules in the interstellar medium, comets, and meteorites: a voyage from dark clouds to the early earth. Ann. Rev. Astron. Astrophys. 38: 427-483.

Emsley, J. 2001. *Nature's Building Blocks:  An A-Z Guide to the Elements*.  Oxford:  Oxford University Press.

Emsley, J. 1998. *The Elements*, 3rd edition. Oxford: Clarendon Press.

Houdek, G. and Gough, D. O. 2011. On the seizmic age and heavy element abundance of the Sun. Monthly Notices of the Royal Astronomical Society (MNRAS). 418(2):1217-1230.

Hoyle, F., Dunbar, D. N. F., Wenzel, W. A. and Whaling, W. 1953. A State in $^{12}$C Predicted from Astronomical Evidence. Physical Review 92: 1095c.

Jarosik, N. et al. 2011. Seven-year Wilkinson Microwave Anisotropy Probe (WMAP) observations: sky maps, systematic errors, and basic results. Astrophysical Journal Supplement Series, 192(2): 1-15.

Jukes, T. H., 1974. Sea-water and the origins of life. Icarus 21:516-517.

Kobayashi, K. and Ponnamperuma, C. 1985. Trace Elements in Chemical Evolution, I. Origins of Life 16: 41-55.

Kragh, H. 2010. When is a prediction anthropic? Fred Hoyle and the 7.65 MeV carbon resonance. Preprint, http://philsci-archive.pitt.edu/id/eprint/5332, Date deposited: 04 May 2010.

Kwok, S. 2004. The synthesis of organic and inorganic compounds in evolved stars. Nature 430: 985-99l.

Lineweaver, C. H., Fenner, Y. Gibson, Y. K. 2004. The Galactic Habitable Zone and the Age Distribution of Complex Life in the Milky Way. Science 02 Jan 2004 : 59-62.

Lodders, K., Palme H., & Gail, H.P. 2009. Abundances of the elements in the solar system. In Landolt Börnstein, New Series, Vol. VI/4B, Chap. 4.4, J.E.Trümper (ed.). Berlin, Heidelberg, New York: Springer Verlag, p. 560-630.

Lodders, K. 2010. Solar system abundances of the elements. Lecture Notes of the Kodai School on «Synthesis of Elements in Stars» held at Kodaikanal Observatory, India, Asprtil 29 – May 13. 2008 (A. Goswami and B.E.Reddy eds.). Astrophysics and Space Science Proceedings, Springer-Verlag Berlin, Heidelberg, p. 379-417.

Lucantoni, M., Luisi, P. L. 2012. On the Universality of the Living: A Few Epistemological Notes. Orig. Life Evol. Biosph. 42: 385-387.

Marboeuf, U., Thiabaud, A., Alibert, Y., Cabral, N. and Benz, W. 2014. From planetesimals to planets: volatile molecules. Astronomy and Astrophysics 570: A36.

Mattson, L. 2009. On the Existence of a Galactic Habitable Zone and the Origin of Carbon. Swedish Astrobiology Meeting, Lund. http://videos.nordita.org/conference/SwAN2009/Mattsson.pdf.

McCabe, M. and Lucas, H. 2010. On the Origin and Evolution of Life in the Galaxy. http://arxiv.org/ftp/arxiv/papers/1104/1104.4322.pdf; also International Journal of Astrobiology 9(04) 217-226.

Orgel, L. E., 1974. Reply: "Comments on 'directed panspermis'" and "seawater and the origin of life". Icarus, 21:518.

Pettini, M. 2003. Element Abundances through the Cosmic Ages. Lectures given at the XIII Canary Islands Winter School of Astrophysics "Cosmochemistry: The Melting Pot of Elements". http://www.ast.cam.ac.uk/pettini/canaries13. arXiv:astro-ph/0303272.

Prantzos, N. 2008. On the "galactic habitable zone". Space Sci. Rev. 135: 313-322.

Salpeter, E. E. 1952. Nuclear reactions in stars without hydrogen. Astrophysical Journal 115: 326-328.

Schroeder, C. 2015. Explainer: what is interplanetary dust and can it spread the ingredients of life? University of Stirling. Published on http://the conversation.com. Posted Nov.13.2015.

Sharov, A. A. 2006. Genome increase as a clock for the origin and evolution of life. Biology Direct, 1: 17.

Sharov, A. A. and Gordon, R. 2013. Life before Earth. arXiv:1304.3381 [physics.gen-ph], Cornell University Library

Sharov, A. A. 2012. A Short Course on Biosemiotics: 2. Evolution of Natural Agents: Preservation, Development, and Emergence of Functional Information. Second Life®: Embryo Physics Course: http://embryogenesisexplained.com/2012/04/a-short-course-on-biosemiotics-2.html

Sneden, C., Cowan, J.J., & Gallino, R. 2008. Neutron-capture elements in the early galaxy. Annu. Rev. Astron. Astrophys. 46, 241-288.

Valković, V. 1990. Origin of Trace Element requirement by Living Matter. Symmetries in Science IV, Eds. B.Gruber and J.H.Yopp, Plenum Press, New York, pp 213-242.

van Dishoeck, E. F. and Blake, G. 1998 Chemical evolution of star-forming regions. Ann. Rev. Astron. Astrophys. 36:317-368.

von Weizsäcker, C. F. 1938. Über Elementumwandlungen in Innern der Sterne II. Physikalische Zeitschrift 39: 633-645.

Wallerstein, G., Iben, I., Parker, P., Boesgaard, A.M., Hale, G.M., Champagne, A.E., Barnes, C.A., Käppeler, F., Smith, V.V., Hoffman, R.D., Timmes, F.X., Sneden, C., Boyd, R.N., Meyer, B.S., & Lambert, D.L. 1997. Synthesis of the elements in stars: forty years of progress. Rev. Mod. Phys. 69, 995-1084.

Wallis, M.K., Wickramasinghe, N.C. 2004. Interstellar transfer of planetary microbiota. Mon. Not. R. Astron. Soc., 348: 52-61.

Woosley, S.E., Heger, A., Weaver, T.A. 2002. The evolution and explosion of massive stars Rev. Mod. Phys. 74, 1015-1072.

Zheng, W. et al. 2012. A magnified young galaxy from about 500 million years after the Big Bang. Nature, 489(7416): 406-408.

**Tables:**

Table 1: Basic nuclear reaction links.

|  | N-2 | N-1 | N | N+1 | N+2 |
|---|---|---|---|---|---|
| **Z+2** |  |  |  | (α,n) | (α,γ) |
| **Z+1** |  | (p,n) | (p,γ) |  | (α,p) |
| **Z** |  | (γ,n) | **[Z,N]** | (n,γ) |  |
| **Z-1** | (p,α) |  | (γ,p) | (n,p) |  |
| **Z-2** | (γ,α) | (n,α) |  |  |  |

Table 2. Dietary elements

| H |  |  |  |  |  |  |  |  |  |  |  |  |  |  |  |  | He |
|---|---|---|---|---|---|---|---|---|---|---|---|---|---|---|---|---|---|
| Li | Be |  |  |  |  |  |  |  |  |  |  | B | C | N | O | F | Ne |
| Na | Mg |  |  |  |  |  |  |  |  |  |  | Al | Si | P | S | Cl | Ar |
| K | Ca |  | Sc | Ti | V | Cr | Mn | Fe | Co | Ni | Cu | Zn | Ga | Ge | As | Se | Br | Kr |
| Rb | Sr |  | Y | Zr | Nb | Mo | Tc | Ru | Rh | Pd | Ag | Cd | In | Sn | Sb | Te | I | Xe |
| Cs | Ba | * | Lu | Hf | Ta | W | Re | Os | Ir | Pt | Au | Hg | Tl | Pb | Bi | Po | At | Rn |
| Fr | Ra | ** | Lr | Rf | Dd | Sg | Bh | Hs | Mt | Ds | Rg | Cn | U3 | Fl | U5 | Lv | U7 | U8 |

**\***Lanthanides: La, Ce, Pr, Nd, Pm, Sm, Eu, Gd, Tb, Dy, Ho, Er, Tm, Yb.
**\*\***Actinides: Ac, Th, Pa, U, Np, Pu, Am, Cm, Bk, Cf, Es, Fm, Md, No.
U3→Ununtrium, Z=113, temporary symbol Uut. U5→Ununpentium, Z=115, temporary symbol Uup. U7→Ununseptium, Z=117, temporary symbol Uus. U8→Ununoctium, Z= 118, temporary symbol Uuo.

Table 3: Abundance ratios.

| Abundance ratios | Values | Ratio |
|---|---|---|
| $(C/O)_{univ.}/(C/O)_{life}$ | 0.5/0.37 | 1.35 |
| $(C/O)_{solar}/(C/O)_{life}$ | 0.54/ 0.37 | 1.46 |
| $(C/O)_{crust}/(C/O)_{life}$ | 0.001/0.37 | 0.0027 |
| $(C/O)_{seawater}/(C/O)_{life}$ | $16 \times 10^{-8}/0.37$ | $43 \times 10^{-8}$ |
| $\Sigma(H,C,N,O)_{univ.}/ \Sigma(H,C,N,O)_{life}$ | 76.6/96.857 | 0.791 |
| $\Sigma(H,C,N,O)_{crust}/ \Sigma(H,C,N,O)_{life}$ | 47.6025/96.857 | 0.491 |
| $\Sigma(H,C,N,O)_{seawater}/ \Sigma(H,C,N,O)_{life}$ | 99.305/96.857 | 1.025 |
| $\Sigma(Na, Mg, P, S, Cl, K, Ca)_{univ.}/ \Sigma(Na, Mg, P, S, Cl, K, Ca)_{life}$ | 0.1201/3.5214 | 0.034 |
| $\Sigma(Na, Mg, P, S, Cl, K, Ca)_{seawater.}/ \Sigma(Na, Mg, P, S, Cl, K, Ca)_{life}$ | 3.3197/3.5214 | 0.943 |
| $\Sigma(Cr, Mn, Fe, Co, Ni, Cu, Zn, Se, Mo, I)_{univ.}/ \Sigma(Cr, Mn, Fe, Co, Ni, Cu, Zn, Se, Mo, I)_{life}$ | 0.1186396/0.009 | 13.18 |
| $\Sigma(Cr, Mn, Fe, Co, Ni, Cu, Zn, Se, Mo, I)_{seawater}/ \Sigma(Cr, Mn, Fe, Co, Ni, Cu, Zn, Se, Mo, I)_{life}$ | 0,09179/0.009 | 10.20 |
| $\Sigma(Li, B, F, Si, V, As)_{univ.}/ \Sigma(Li, B, F, Si, V, As)_{life}$ | 0.0701414/0.0079 | 7.793 |
| $\Sigma(Li, B, F, Si, V, As)_{seawater}/ \Sigma(Li, B, F, Si, V, As)_{life}$ | 0.00205245/0.0079 | 0.2598 |

Table 4: History of Universe milestones (scale in $10^9$ years).

| Time/$10^9$ years | Event | Ref. |
|---|---|---|
| -13.75 | Big Bang | Jarosik et al. 2011. |
| -13.55 | Dark ages ending | Zheng et al. 2012. |
| -13.2 | EGSY8p7 Galaxy | Zitrin et al. 2015. |
| -11.3+6.7 | Origin of life | This work |
| $-9.7 \pm 2.5$ | Origin of DNA | Sharov 2006; 2012. |
| $-8.5 \pm 5$ | Start of Galactic habitability | McCabe and Lucas 2010. |
| -4.57 | Sun forms | Christensen-Dalsgaard et al. 1996. |
| -4.56 | Earth forms | Houdek and Gough 2011 |
| 0 | Now | |

**Figure Captions:**

Fig. 1. Living matter: concentrations of chemical elements.

**Figures**

Fig. 1:

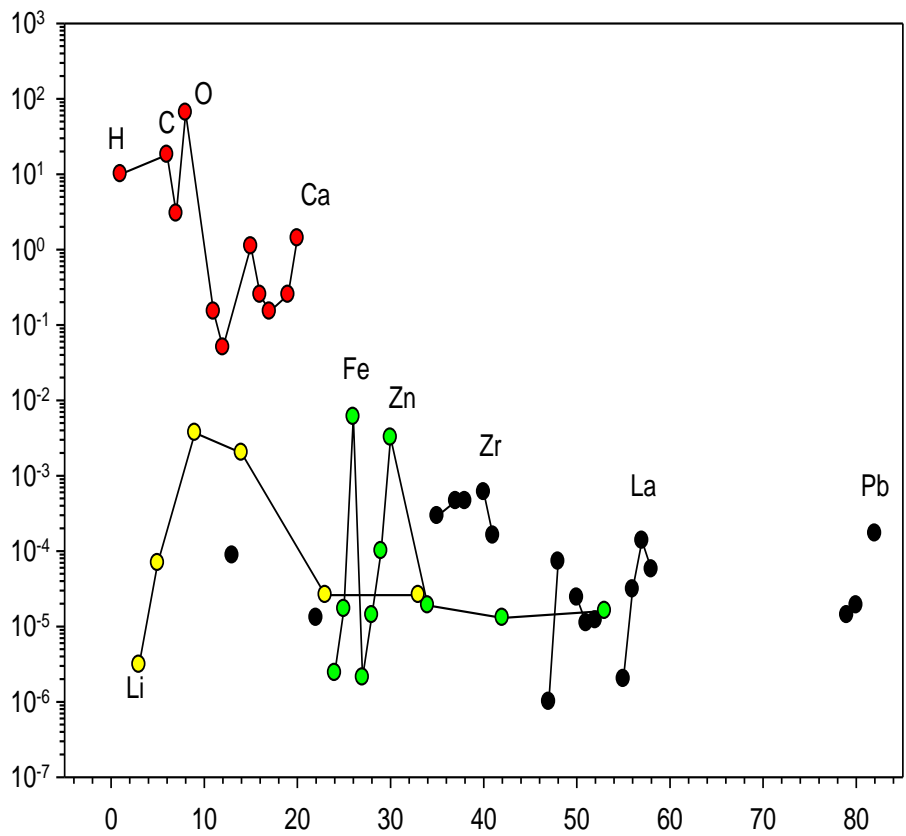